\documentclass[letterpaper]{article} 
\usepackage{aaai2026}  
\usepackage{times}  
\usepackage{helvet}  
\usepackage{courier}  
\usepackage[hyphens]{url}  
\usepackage{graphicx} 
\usepackage{natbib}  
\usepackage{caption} 
\usepackage{algorithm}
\usepackage{algorithmic}

\usepackage{newfloat}
\usepackage{listings}
\DeclareCaptionStyle{ruled}{labelfont=normalfont,labelsep=colon,strut=off} 
\floatstyle{ruled}
\newfloat{listing}{tb}{lst}{}
\floatname{listing}{Listing}
\title{Learning Conjugate Direction Fields for Planar Quadrilateral Mesh Generation}

\author{
    Jiong Tao\textsuperscript{\rm 1}, 
    Yong-Liang Yang\textsuperscript{\rm 1}, 
    Bailin Deng\textsuperscript{\rm 2}\thanks{Corresponding author.}
}

\affiliations{
    \textsuperscript{\rm 1} Department of Computer Science, University of Bath, United Kingdom \\ 
    \textsuperscript{\rm 2} School of Computer Science and Informatics, Cardiff University, United Kingdom \\
    jt2337@bath.ac.uk, y.yang@cs.bath.ac.uk, DengB3@cardiff.ac.uk
}

\usepackage{bibentry}

\usepackage{amsmath}
\usepackage{multirow}
\usepackage{amssymb}

\newcommand{\vnormal}{\mathbf{n}}
\newcommand{\vpos}{\mathbf{p}}
\newcommand{\vprojvec}{\mathbf{l}}
\begin{document}

\maketitle

\begin{abstract}
Planar quadrilateral (PQ) mesh generation is a key process in computer-aided design, particularly for architectural applications where the goal is to discretize a freeform surface using planar quad faces. The conjugate direction field (CDF) defined on the freeform surface plays a significant role in generating a PQ mesh, as it largely determines the PQ mesh layout. 
Conventionally, a CDF is obtained by solving a complex non-linear optimization problem that incorporates user preferences, i.e., aligning the CDF with user-specified strokes on the surface. This often requires a large number of iterations that are computationally expensive, preventing the interactive CDF design process for a desirable PQ mesh. 
To address this challenge, we propose a data-driven approach based on neural networks for controlled CDF generation. Our approach can effectively learn and fuse features from the freeform surface and the user strokes, and efficiently generate quality CDF respecting user guidance.
To enable training and testing, we also present a dataset composed of 50000+ freeform surfaces with ground-truth CDFs, as well as a set of metrics for quantitative evaluation.
The effectiveness and efficiency of our work are demonstrated by extensive experiments using testing data, architectural surfaces, and general 3D shapes.

\end{abstract}


\begin{links}
    \link{Code}{https://github.com/jiongtj/Learning-CDF}
\end{links}


\section{Introduction}

Planar quadrilateral (PQ) meshes are special piecewise linear surface representations that consist entirely of planar quad faces. They are widely used in computer-aided design, particularly for modeling discrete architectural surfaces, due to several key advantages~\cite{Pottmann2013,POTTMANN2015145}.
First, the planarity of quad faces significantly reduces the cost of fabricating panels with physical materials such as glass.
Second, compared to common triangle meshes, PQ meshes have lower vertex valence, resulting in reduced complexity for conjoining supporting beams along each edge. 
Third, the edges of PQ meshes often form intuitive layouts that align with human aesthetic preferences.

Given a freeform surface representing the desired design geometry, discretizing it into a PQ mesh is a challenging problem that has been studied extensively~\cite{liu2006geometric,zadravec2010designing,liu2011general}.
The process generally consists of two main stages.
In the first stage, an initial quad mesh with a plausible distribution of elements (vertices, edges, and faces) is generated to serve as the mesh layout. 
In the second stage, a geometric optimization is performed to refine the layout and enforce planarity of each quad face by modifying vertex positions. 
Due to the non-convex nature of the optimization problem, a key challenge is ensuring a high-quality initial layout, such that the subsequent optimization can successfully generate a PQ mesh that closely approximates the input freeform surface.

Prior works control the edge distribution of the initial mesh layout using a conjugate direction field (CDF) such as the principal (curvature) direction field (PDF)~\cite{liu2006geometric}, where the mesh edges align with two families of polylines that are discrete counterparts of integral curves of the CDF. The conjugacy of the CDF ensures the initial mesh faces are approximately planar, which is crucial for the success of subsequent PQ mesh optimization~\cite{liu2006geometric}.
However, unlike PDFs which are uniquely determined by surface geometry except in isotropic regions, CDFs are not unique and have high degrees of freedom over the surface. 
Therefore, additional conditions are needed to 
derive a CDF that is consistent with the user's design intent. 
A common approach is to introduce user-specified strokes on some parts of the surface to indicate preferred CDF directions.
These strokes serve as constraints for a non-linear optimization that computes a smooth CDF suitable for quad mesh initialization~\cite{liu2011general}. 
However, this process is highly time-consuming, often requiring many iterations of user input and optimization to obtain a satisfactory CDF.

To improve the efficiency of CDF design, we propose a learning-based approach that can directly generate CDFs without complex optimization. 
Instead of hand-crafting features, we use neural networks to learn surface features from 
geometric information such as vertex positions and normals.
Our approach supports conditional field generation, where user-provided strokes guide the CDF computation. This capability enables generating diverse fields with different flow behaviors on the same surface, which is valuable for design exploration.
To provide meaningful control, we propose a global stroke representation that can be seamlessly integrated with the surface representation for feature extraction.
Since collecting real-world data from actual design sessions is impractical, we construct a synthetic dataset inspired by~\cite{deng2022sketch2pq}. The synthesis process emulates real design workflows by tracing streamlines of a ground truth CDF to approximate user strokes. 
This dataset enables comprehensive training and testing of our learning-based CDF generation approach.
Extensive experiments demonstrate that our method can effectively generate CDFs that respect user constraints, while being an order of magnitude faster than traditional optimization techniques. Moreover, it generalizes well to real-world architectural surfaces and general 3D shapes, making it well-suited for practical design applications.

In summary, our main contributions are:
\begin{itemize}

    \item A novel approach for efficient CDF generation using deep neural networks, avoiding complex optimization.
   
    \item A conditional CDF generation method that incorporates user-specified stroke constraints based on a novel global stroke representation.
 
    \item A large-scale synthetic dataset that enables training and testing for data-driven CDF generation.
    
\end{itemize}

\section{Related Work}

\paragraph{PQ Mesh}
Planar quadrilateral (PQ) meshes, where all faces are quadrilaterals with coplanar vertices, 
are commonly used in architectural design and fabrication ~\cite{glymph2004parametric, pottmann2007geometry}, particularly for glass structures.  
Unlike general quadrilateral meshes that can be structured in different ways~\cite{bommes2013} (e.g., using Morse-Smale complexes~\cite{Dong2006, Huang2008}, scalar fields~\cite{DONG2005, Tong2006}, vector/cross fields~\cite{Alliez2003, Ray2006, Kaelberer2007, bommes2009mixed, ebke2013qex, Lyon2019}, hybrid representations~\cite{Fang2018}), PQ meshes are a discrete counterpart of conjugate curve networks on surfaces~\cite{Bobenko2006}.
~\cite{liu2006geometric} derived quad mesh along the principal curvature lines and applied vertex perturbation optimization to achieve face planarity. ~\cite{zadravec2010designing} formulated the design of conjugate curve networks as a vector field design problem, and obtained CDF by optimizing the vector field. Later, ~\cite{liu2011general} extracted general PQ meshes by computing smooth CDF on reference surfaces, satisfying user-defined directional constraints. ~\cite{diamanti2014designing} utilized a $2^2$ Polyvector field as initialization, deforming it to the closest CDF. However, the aforementioned methods mainly formulated the CDF design as a non-linear optimization problem involving numerous variables and a set of non-linear constraints. Due to its non-linearity, solving such an optimization problem is highly challenging. Moreover, as the model size increases, the computational cost becomes prohibitively high for interactive applications. 
Recently, a novel learning-based method for PQ mesh design using a single sketch has been proposed by~\cite{deng2022sketch2pq}. It allows users to draw structural lines as the surface boundary and contours, annotate occlusions, and define sparse feature lines indicating the directions of PQ mesh edges. 
During training, labeled PQ mesh data is required as ground truth, making it challenging to generate a sufficient number of PQ meshes for constructing a comprehensive training dataset. Moreover, this method takes a single-view 2D sketch as input and only predicts projected edge directions in 2D, which limits its usage for intuitive CDF generation on 3D freeform surfaces and general shapes.

\paragraph{Deep Learning on 3D Data}
In recent years, various deep learning methods on 3D data have been proposed to address 3D vision and graphics problems, such as point cloud processing~\cite{qi2017pointnet, wang2019dynamic}, connectivity construction from discrete data~\cite{sharp2020pointtrinet, rakotosaona2021differentiable}, neural mesh processing~\cite{potamias2022neural, chen2023neural}, 3D shape generation~\cite{hui2022neural, zhang20233dshape2vecset}, just to name a few. A comprehensive review is beyond the scope of this paper and more background information can be found in related surveys~\cite{bronstein2017geometric, guo2020deep, xiao2020survey, xu2023survey}.
In our work, we apply the widely used DGCNN~\cite{wang2019dynamic} to extract both local and global latent features from the input shape while seamlessly fusing features respecting the stroke guidance.

\paragraph{Learning Directional Fields}
With the advances of 3D deep learning, the generation of directional fields based on neural networks has also attracted research attention. ~\cite{girard2020regularized} learned a frame field from satellite images to align with object boundaries, which improves segmentation accuracy. ~\cite{dielen2021learning} proposed a learning-based frame fields generation method for field-guided quadrangulation. However, this method is trained and tested merely on a human body dataset which highly restricts its generalization capabilities for freeform PQ mesh generation. Also, the frame field is directly predicted based on the geometric shape (i.e., the human body) without user guidance, thus unsuitable for interactive architectural design purposes. 
A self-supervised cross field learning framework was developed by~\cite{wei2023ipunet} for feature-aligned point cloud unsampling. A novel neural framework for learning vector/cross fields, based on vector heat diffusion, was designed by~\cite{gao2024vectorheatnet}, offering the invariance to rigid transformation and isometric deformation of the inputs. \cite{Dong2025NeurCross} introduced a self-supervised cross field generation method, exploring the connections between the neural representation of a triangle mesh and its PDF for quadrangulation. 
The cross fields learned by these methods stem from PDFs, and thus cannot be generalized for CDF-based PQ mesh generation as in our work.

\section{Preliminary}

Direction fields on surfaces are fundamental to quadrilateral mesh generation, as they can guide the orientation and flow of mesh edges. Among them, Conjugate Direction Fields (CDF) provide a compact representation that encodes the locally planar structures essential for generating Planar Quadrilateral (PQ) meshes.

We first formally define a CDF. On a triangle mesh
$\mathcal{M}=\{\mathcal{V}, \mathcal{F} \}$, a pair of direction fields  $\{(\mathbf{u}_j, \mathbf{v}_j)\}$ defined on each mesh face $f_j$ is called \emph{conjugate} if the following condition is satisfied:
\begin{equation}
    \kappa_{j,1} (\mathbf{u}_j \cdot \mathbf{d}_{j,1}) (\mathbf{v}_j \cdot \mathbf{d}_{j, 1}) +  \kappa_{j,2} (\mathbf{u}_j \cdot \mathbf{d}_{j,2}) (\mathbf{v}_j \cdot \mathbf{d}_{j,2}) = 0,
\label{eq:conjugacy}
\end{equation}
where $\mathbf{d}_{j, 1}, \mathbf{d}_{j, 2}$ are  unit principal direction vectors on $f_j$, and
$\kappa_{j, 1}, \kappa_{j, 2}$ are the corresponding principal curvatures. 

The Principal Direction Field (PDF), which consists of the principal direction pairs $(\mathbf{d}_{j, 1}, \mathbf{d}_{j, 2})$ over the surface, is a special and well-known case of a CDF, as it trivially satisfies the conjugacy condition~\eqref{eq:conjugacy}.
However, unlike a PDF, which is uniquely determined by surface geometry (except at umbilic points), a general CDF is not unique. This non-uniqueness provides high degrees of freedom, offering significant design flexibility for PQ mesh generation.

This flexibility, however, also presents a significant challenge: a CDF is ill-defined by geometry alone. To specify a single, desirable CDF from this high-dimensional space, additional conditions are required. Typical conditions include:
\begin{enumerate}
    \item \emph{Field Smoothness}: To ensure a regular and coherent mesh layout while avoiding unnecessary singularities, the field must be smooth across neighboring faces.
    \item \emph{User-Defined Constraints}: To capture design intent, the field should adhere to various user-specified conditions. One primary example is the alignment with user-provided strokes on the surface.
\end{enumerate}

Traditional methods formulate the generation of a smooth CDF that respects user constraints as a complex, constrained non-linear optimization problem~\cite{liu2011general}:
\begin{eqnarray}
\min_{\mathbf{u}, \mathbf{v}} & & E_s(\mathbf{u}, \mathbf{v}) \nonumber \\
\text{s.t.} & & \text{Conjugacy constraints}, \nonumber \\
& & \text{User-specified constraints}.
\label{eq:energy}
\end{eqnarray}
Here, $E_{s}$ represents a smoothness energy term that measures the variation of the field across the surface.

Due to the non-linearity of this optimization problem (particularly the conjugacy constraint), its solution is highly challenging and computationally expensive. This process often becomes a significant bottleneck, preventing the interactive CDF design and iterative exploration required in practical architectural applications. To address this challenge, this paper proposes a learning-based method to produce high-quality CDFs efficiently, avoiding the costly numerical optimization process entirely.

\begin{figure}[t!]
    \includegraphics[width=\linewidth]{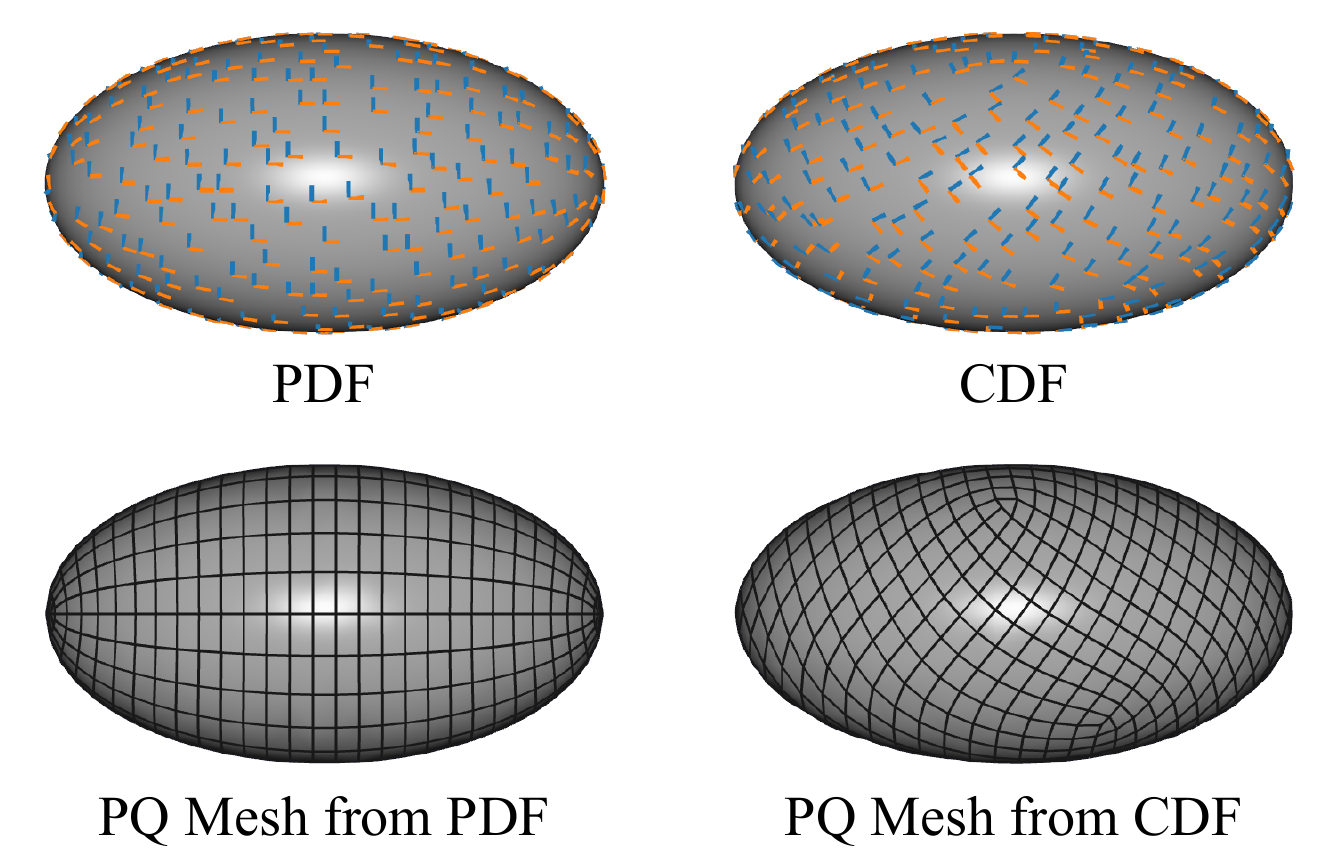}
	\caption{Visualization of direction fields and generated meshes on an ellipsoid.}
\label{fig:Ellipsoid}
\end{figure}

\begin{figure*}[t]
    \includegraphics[width=\textwidth]{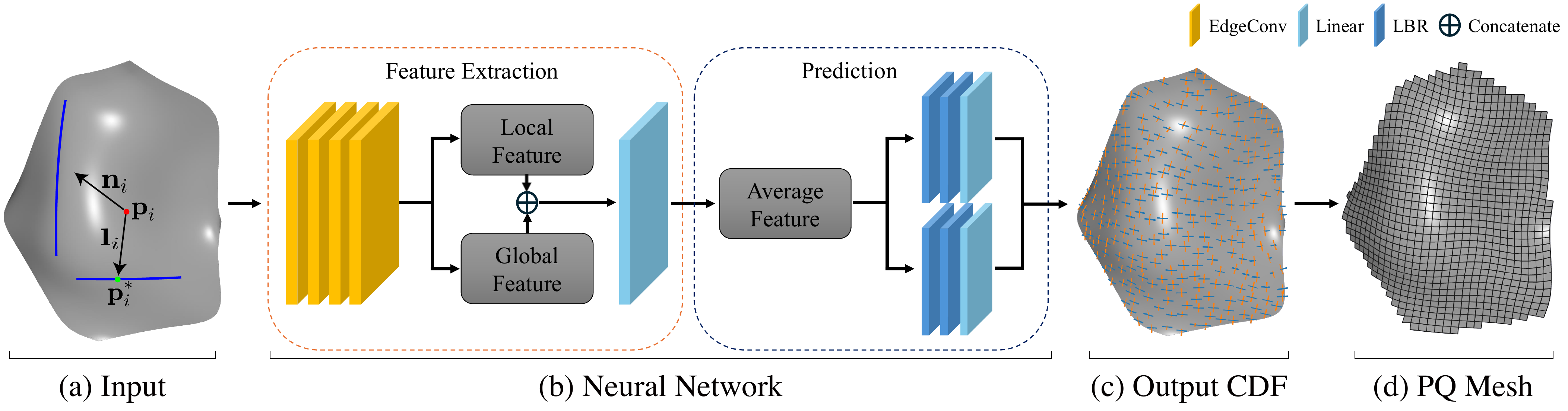}
	\caption{Our pipeline for inferring the conjugate direction field for PQ mesh generation. There are two core components in our learning model, including feature extraction and CDF prediction. In the first component, we take vertex positions, vertex normals, and the projected vectors to the user strokes as input, obtaining the latent feature for each vertex. In the second component, we use two MLPs to infer a CDF over the surface based on the latent features. In the figure, LBR means the combination of Linear, BatchNorm, and ReLU layers.}
\label{fig:pipeline}
\end{figure*}

After obtaining the CDF, PQ mesh generation aims to construct a quad mesh whose edges are aligned with the optimized field. Generally, an initial quad mesh is generated via a global parameterization method from CDFs and then refined through numerical optimization to produce high-quality, planar quads~\cite{liu2011general}, as illustrated in Fig.~\ref{fig:Ellipsoid}.

%

\section{Method}

\subsection{Overview}
The pipeline of our learning-based method is shown in Fig.~\ref{fig:pipeline}.
Given a triangle mesh $\mathcal{M}=\{\mathcal{V}, \mathcal{F} \}$ representing the reference freeform surface and a set of user-specified stroke curves $\mathcal{S} = \{ \mathbf{S}_i \}$ on the mesh, our goal is to learn two families of (piecewise-linear) direction fields $\{(\mathbf{u}_j, \mathbf{v}_j)\}$ defined on each mesh face $f_j$.
These direction fields should satisfy two properties: i) the local conjugacy to form a valid CDF, and ii) the alignment with the user-specified stroke curves to reflect design intent.
%
%
Once the CDF is computed, we derive a global parameterization whose iso-parameter lines align with the CDF~\cite{bommes2009mixed}, and then extract a quad mesh from this parameterization~\cite{ebke2013qex} with mesh edges forming a discrete counterpart of the conjugate curve network. The initial quad mesh is further optimized by perturbing vertex positions to enforce quad face planarities~\cite{deuss2015shapeop}, resulting in the final PQ mesh.

The key advantages of our learning-based approach are twofold. First, by avoiding costly nonlinear optimization, it significantly reduces computation time for CDF generation. Second, by taking user strokes as input, it empowers designers with control over the CDF and, consequently, the PQ mesh structure. This is valuable for design exploration and for generating PQ meshes aligned with design intent.
The rest of the subsections elaborate on our feature representation, network architecture, and loss functions, respectively.

\subsection{Feature Representation}
\label{sec:FeatureRep}
To accurately predict the CDF over the surface while respecting user-specified stroke constraints, we must carefully design the input features for our neural network. The most fundamental feature is the vertex position $\vpos_i \in \mathbb{R}^3$, which captures the geometry of the mesh.

Besides vertex positions, we also include vertex normals $\vnormal_i \in \mathbb{R}^3$ as input features. Vertex normals encode local surface orientation and help the network better understand the shape of the surface in the neighborhood of each vertex. By providing both positions and normals, we give the network a more complete characterization of the local geometry.

To represent the user-specified stroke curves, we compute a vertex-wise stroke feature as follows. For each vertex $\vpos_i$, we find its closest point $\vpos_i^\ast$ among all stroke curve points. We then compute a projection vector $\vprojvec_i = \vpos_i^\ast - \vpos_i$ that connects the vertex to its closest point on the stroke curves (see Fig.~\ref{fig:pipeline}a). This projection vector encodes the spatial relationship between the vertex and the stroke curves, providing a meaningful stroke representation at each vertex.

Our final input feature vector for each vertex is a concatenation of the vertex position $\vpos_i$, vertex normal $\vnormal_i$, and stroke projection vector $\vprojvec_i$, resulting in a 9-dimensional feature vector. This compact yet informative feature representation enables our network to learn to generate CDFs that align with user strokes while adhering to the surface geometry.

\subsection{Network Architecture}
\label{sec:Architecture}

Our network architecture, illustrated in Fig.~\ref{fig:pipeline}b, takes the 9-dimensional per-vertex feature vectors as input and predicts the CDF vectors ${(\mathbf{u}_j, \mathbf{v}_j)}$ on each face. The network consists of two main components: a feature extraction module and a direction field prediction module.

\subsubsection{Feature Extraction Module}
Inspired by the method of~\cite{dielen2021learning}, our feature extraction module is designed to capture both local and global geometric information. 
However, instead of using separate networks for local and global features as in~\cite{dielen2021learning}, we employ  a single unified architecture based on the Dynamic Graph CNN (DGCNN)~\cite{wang2019dynamic} 

DGCNN is a powerful graph neural network that operates directly on unstructured point clouds. It extracts features by applying EdgeConv operations on dynamic local graph neighborhoods, which are recomputed at each layer based on the learned point embeddings. This allows the network to learn both local and global shape properties adaptively. DGCNN has achieved strong performance across a range of 3D shape analysis tasks thanks to its ability to capture multi-scale geometric features.

 Our network takes the vertex positions $\{\vpos_i\}$, vertex normals $\{\vnormal_i\}$ and projection vectors $\{\vprojvec_i\}$ as input, and 
 outputs a 256-dimensional latent feature for each vertex. Specifically, following DGCNN~\cite{wang2019dynamic}, we use a sequence of four EdgeConv layers to extract vertex features. However, instead of transforming the vertex features into a global shape feature by applying a fully-connected layer with ReLU activation~\cite{nair2010rectified} and a pooling strategy as in~\cite{wang2019dynamic}, we concatenate the local feature of each vertex with the global shape feature over the entire surface to construct an integrated feature. Finally, a fully-connected layer is applied to the integrated feature to obtain the final feature representation for each vertex. It is important to note that this feature integrates both global shape information and local geometry information.

\subsubsection{Prediction Module}
The direction field prediction module takes the 256-dimensional per-vertex features as input and predicts the CDF vectors on $(\mathbf{u}_j, \mathbf{v}_j)$ on each triangle face. To obtain face features from the vertex features, we simply average the vertex features for each triangle.

We use two separate MLPs to predict the two families of direction vectors $\{\mathbf{u}_j\}$ and $\{\mathbf{v}_j\}$ independently. Each network consists of three layers that map the input 256-dimensional face feature to intermediate representations of dimensions 128 and 64, and finally output a 3-dimensional vector representing either $\mathbf{u}_j$ or $\mathbf{v}_j$. The first two layers use BatchNorm~\cite{ioffe2015batch} and ReLU~\cite{nair2010rectified} as the activation function. Since the last layer directly creates the directional vectors, we remove any BatchNorm and activation function in that layer. Finally, we normalize each predicted vector to unit length.

\subsection{Loss Functions}
\label{sec:Losses}
We employ a combination of loss terms to train our network to generate smooth, locally conjugate direction fields that align with the user-specified strokes.

\subsubsection{Direction Alignment}
\label{sec:DirectionLoss}

Our first requirement is to align the predicted direction field $\{(\mathbf{u}_j, \mathbf{v}_j)\}$ with the ground truth directions $\{(\mathbf{u}_j^{*}, \mathbf{v}_j^{*})\}$. Different from direct computation of the cosine similarity between two fields to measure alignment, as performed in~\cite{dielen2021learning}, we use the following function to measure the alignment between $\mathbf{u}_j$ and $\mathbf{u}_j^{*}$ and between $\mathbf{v}_j$ and $\mathbf{v}_j^{*}$:
\begin{equation}
{E}_j = (\mathbf{u}_j \cdot \mathbf{u}^{* \perp}_j)^2 + (\mathbf{v}_j \cdot \mathbf{v}^{* \perp}_j)^2.
\label{equ:loss1}
\end{equation}
Here $\mathbf{u}^{* \perp}_j$, $\mathbf{v}^{* \perp}_j$ are 90-degree rotations of the ground-truth unit directions $\mathbf{u}^{*}_i$, 
$\mathbf{v}^{*}_i$ about the face normal $\mathbf{n}_j$.

Note that our use of rotated ground truth vectors to measure alignment enables us to handle the sign ambiguity for the ground-truth vectors, i.e., it allows the $\mathbf{u}_j$ to be close to either $\mathbf{u}^{* \perp}_j$ or  $-\mathbf{u}^{* \perp}_j$ (and the same for $\mathbf{v}_j$ and $\mathbf{v}^{* \perp}_j$). This is beneficial as both configurations are feasible for the subsequent parameterization and quad mesh generation. 

However, for more effective learning, we shall not prescribe specific correspondence between the predicted direction vectors $(\mathbf{u}_j, \mathbf{v}_j)$ and the ground truth directions $(\mathbf{u}_j^*, \mathbf{v}_j^*)$. Instead, we only need the predicted pair to align well with the ground truth pair as a whole, regardless of which predicted vector corresponds to which ground truth vector. To handle this ambiguity, we also measure the alignment for the other correspondence with the following term:
\begin{equation}
{E}'_j = (\mathbf{u}_j \cdot \mathbf{v}^{* \perp}_j)^2 + (\mathbf{v}_j \cdot \mathbf{u}^{* \perp}_j)^2.
\end{equation}
We then use the minimum value between ${E}_j$ and ${E}'_j$ to indicate the alignment between the vector pairs $(\mathbf{u}_j, \mathbf{v}_j)$  and $(\mathbf{u}_j^*, \mathbf{v}_j^*)$ as a whole, and introduce the following direction alignment term to measure the alignment between the predicted direction field with the ground truth:
\begin{equation}
    \mathcal{L}_d = \frac{1}{m} \sum\nolimits_{j=1}^m \min (E_j, E'_j),
\end{equation}
where $m$ is the number of mesh faces.

\subsubsection{Direction Consistency with Normals}
The output direction field $\{(\mathbf{u}_j, \mathbf{v}_j)\}$ should be orthogonal to the corresponding unit face normals $\{\mathbf{n}_j\}$. This can be enforced using the following loss term: 
\begin{equation}
    \mathcal{L}_{dn} = \frac{1}{m} \sum\nolimits_{j=1}^m (\mathbf{u}_j \cdot \mathbf{n}_j)^2 + (\mathbf{v}_j \cdot \mathbf{n}_j)^2.
\end{equation}

\subsubsection{Direction Smoothness}
A smooth CDF is essential to avoid unnecessary singularities and ensure a coherent layout for the final PQ mesh.
For a pair of direction vectors $(\mathbf{u}_j, \mathbf{v}_j)$ on face $f_j$, and another pair $(\mathbf{u}_k, \mathbf{v}_k)$ on a neighboring face $f_k$ that share a common edge with $f_j$, the local CDF smoothness between $f_j$ and $f_k$ can be defined as~\cite{jakob2015instant}:
\begin{equation}
{E}_{jk} = (\hat{\mathbf{u}}_j \cdot \mathbf{u}^{\perp}_k)^2 + (\hat{\mathbf{v}}_j \cdot \mathbf{v}^{\perp}_k)^2,
\end{equation}
where $\hat{\mathbf{u}}_j$, $\hat{\mathbf{v}}_j$ are the vectors obtained by parallel transporting $\mathbf{u}_j$, $\mathbf{v}_j$ from face $f_i$ to face $f_k$, respectively~\cite{jakob2015instant}.
$\mathbf{u}^{\perp}_k$, $\mathbf{v}^{\perp}_k$ are the 90-degree rotations of $\mathbf{u}_k$, $\mathbf{v}_k$ about the face normal $\mathbf{n}_k$. 
Here we measure the direction alignment in the same way as Eq.~\ref{equ:loss1}. 

As discussed in Sec.~\ref{sec:DirectionLoss}, the correspondence between direction vectors $(\mathbf{u}_j, \mathbf{v}_j)$ and $(\mathbf{u}_k, \mathbf{v}_k)$ on neighboring faces is ambiguous. Therefore, we also define the smoothness measure for the other correspondence as follows:
\begin{equation}
{E}'_{jk} = (\hat{\mathbf{u}}_j \cdot \mathbf{v}^{\perp}_k)^2 + (\hat{\mathbf{v}}_j \cdot \mathbf{u}^{\perp}_k)^2.
\end{equation}
Finally, the overall direction smoothness of the CDF on the entire mesh is defined as:
\begin{equation}
\mathcal{L}_{ds} = \frac{1}{|\mathcal{N}|}\sum\nolimits_{(j,k) \in \mathcal{N}} \min(E_{jk}, E'_{jk}),
\end{equation}
where $\mathcal{N}$ is the index set for neighboring faces.

\subsubsection{Direction Consistency with Strokes}
To ensure that the inferred CDF closely aligns with the user-specified strokes, we introduce an additional consistency loss between the output direction vectors and the strokes.
For a pair of predicted directions $(\mathbf{u}_k, \mathbf{v}_k)$ on face $f_k$ which contains a segment $\mathbf{s}_k$ (represented as a vector) of a stroke $\mathbf{S}_i$, one of these two directions should align with the segment. 
Following the alignment measurement in Eq.~\ref{equ:loss1}, this can be expressed as:
\begin{equation}
D_{k} = 
\left.{\min((\mathbf{u}_k \cdot \mathbf{s}^{\perp}_k)^2, (\mathbf{v}_k \cdot \mathbf{s}^{\perp}_k)^2)}\right/{\|\mathbf{s}_k\|},
\end{equation}
where $\mathbf{s}^{\perp}_k$ is 90-degree rotation of the stroke segment $\mathbf{s}_k$ about the face normal $\mathbf{n}_k$.
The overall direction consistency with strokes can be written as:
\begin{equation}
    \mathcal{L}_{dc} = \frac{1}{|\mathcal{S}|} \sum\nolimits_{\mathbf{S}_i \in \mathcal{S}} \frac{1}{|\mathcal{T}_i|}\sum\nolimits_{k \in \mathcal{T}_i} D_k,
\end{equation}
where $\mathcal{T}_i$ denotes the index set of faces that contain stroke segments of $\mathbf{S}_i$.

\subsubsection{Field Regularization}
To avoid obtaining a trivial zero direction field, we also introduce a regularization loss:
\begin{equation}
    \mathcal{L}_{fr} = \frac{1}{m}\sum\nolimits_{j=1}^m (\| \mathbf{u}_j \| - 1 )^2 + (\| \mathbf{v}_j \| -1 )^2.
\end{equation}

\subsubsection{Total Loss}
Overall the full loss for our model  is given by:
\begin{equation}
    \mathcal{L}_{total} = \mathcal{L}_d + \lambda_1 \mathcal{L}_{dn} + \lambda_2 \mathcal{L}_{ds} + \lambda_3 \mathcal{L}_{dc} + \lambda_4 \mathcal{L}_{fr},
\end{equation}
where $\lambda_1, \lambda_2, \lambda_3, \lambda_4$ are the loss weights.

\begin{figure*}[thb]
    \centering
    \includegraphics[width=\textwidth]{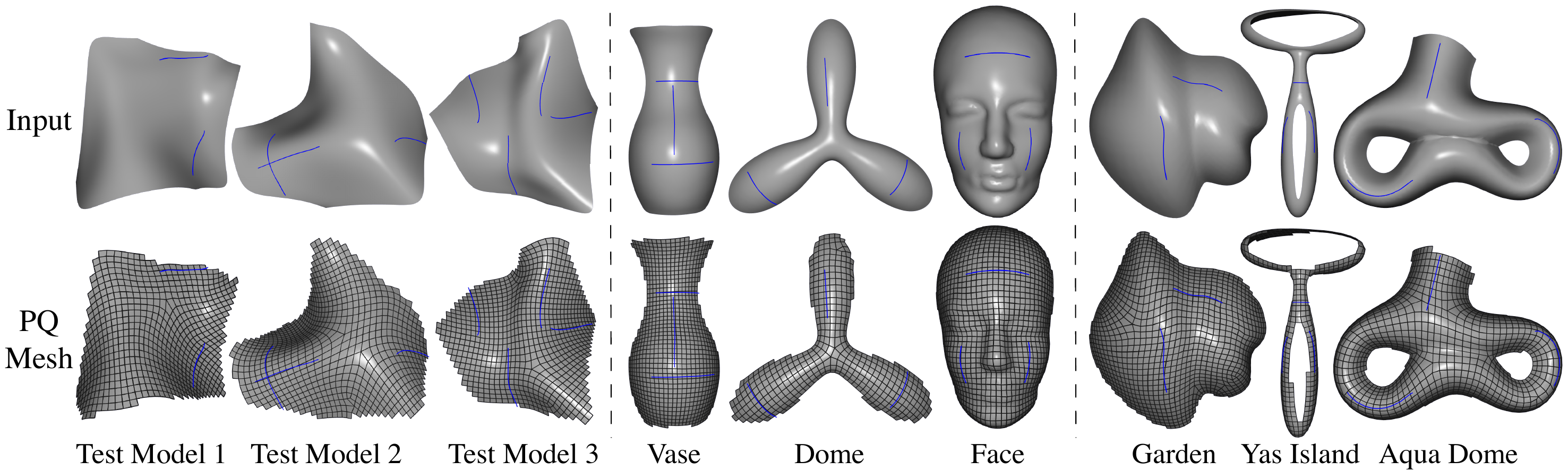}
	\caption{PQ meshes generated using conjugate direction field inferred by our network on different types of surfaces as input. The blue lines on the surface are input strokes, which are streamlines traced from ground-truth CDF for comparison purposes. Left: B-spline surfaces from our test set; Middle: Open-boundary surfaces; Right: Real architectural surfaces. The input triangle meshes and strokes, as well as the output PQ meshes, are provided in the supplementary code and data.}
\label{fig:Viewer_example}
\end{figure*}

\section{Dataset}
\label{sec:Dataset}

To facilitate the development of learning-based approaches for CDF generation on 3D surfaces, we construct a new dataset consisting of freeform triangle mesh surfaces, ground truth CDFs, and corresponding guiding strokes.

\subsection{Freeform Surface Generation}
Following the approach in~\cite{deng2022sketch2pq}, we generate freeform shapes using B-splines. 
The control points of the B-spline surfaces are distributed such that the resulting shapes resemble height-field-like patches. We also apply random 3D deformations to introduce more shape variations.
B-splines are advantageous for representing smooth freeform shapes, particularly architectural surfaces, and allow for easy shape control through manipulating control points.

After obtaining the B-spline surface, we further construct a triangle mesh as its discrete counterpart by evenly sampling 2601 points with 5000 faces. Our dataset comprises 50000, 2500, and 300 freeform triangle mesh surfaces for training, validation, and testing, respectively.

\subsection{Data Normalization}
To facilitate the learning process, we normalize the position, orientation, and size of each triangle mesh as follows.
Given the vertex coordinates $\mathcal{V}= \{\vpos_i \in \mathbb{R}^3 \}$ of the surface mesh $\mathcal{M} =\{\mathcal{V}, \mathcal{F}\}$, we perform Principal Component Analysis (PCA) to estimate the three principal directions $\mathbf{d}_1$, $\mathbf{d}_2$, $\mathbf{d}_3$ based on their corresponding eigenvalues in descending order. We then rotate the mesh to align the principal directions with the global  $x$-, $y$-, and $z$-axes, respectively. Finally, we translate and scale the mesh to fit within a unit sphere.

\subsection{CDF Generation and Stroke Mimicking}
\label{seq:CDFGen}
To obtain diverse ground truth CDFs while simulating user-specified strokes as design guidance, we first randomly sample 1 to 5 anchor points on each triangle mesh. Then we initialize conjugate direction vectors at each anchor point by randomly assigning one direction and computing its conjugate. 
Subsequently, we apply the CDF optimization method from~\cite{diamanti2014designing}, as implemented in~\cite{Directional}, to generate a dense ground truth CDF subject to the direction constraints at the anchor points. 
To mimic user-defined strokes in practical design scenarios, we trace streamlines of the ground truth CDF starting from the anchor points.
The visualization of some representative results is presented in the supplementary materials. 

\section{Experiments}

\begin{figure}[t]
    \includegraphics[width=\linewidth]{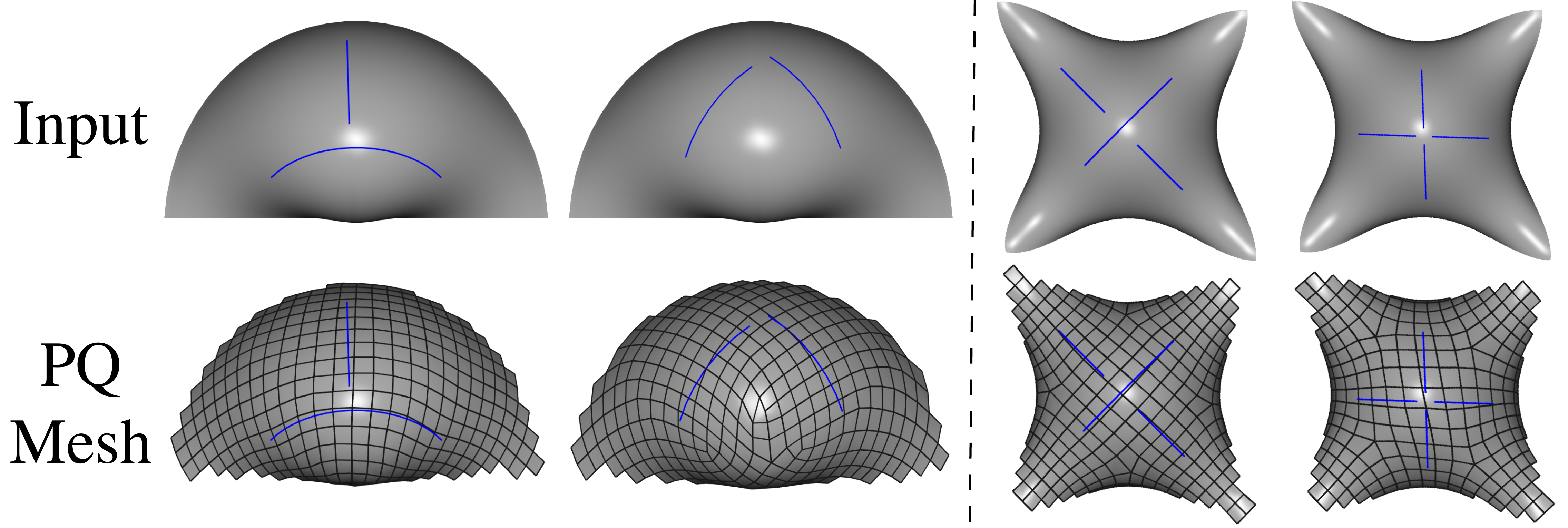}
	\caption{The examples where the different PQ mesh layouts are generated on the same surface with two different types of strokes as inputs.}
\label{fig:Layout_example_2}
\end{figure}

In our experiments, we train the network described in Sec.~\ref{sec:Architecture} for 200 epochs, using 50000 training examples from the dataset generated using the approach described in Sec.~\ref{sec:Dataset}. Throughout the training phase, we utilize the Adam optimizer~\cite{kingma2014adam} with a fixed learning rate of $1.0 \times 10^{-4}$. For the loss weights, we set $\lambda_1 = \lambda_2 = \lambda_3 = \lambda_4 = 1.0$.
To validate the effectiveness and efficiency of our work, we conduct a set of evaluations on a desktop PC with a 24-core Intel Core i9-14900K at 3.20GHz, and an NVIDIA GeForce RTX 4090 GPU with 24GB of video memory.

\subsection{Results}
We test our method on a diverse set of surfaces, including B-spline surfaces from our test set, general open-boundary surfaces, and real architectural surfaces. Fig.~\ref{fig:Viewer_example} demonstrates the PQ meshes generated using our method on these different surface types. 
Moreover, benefiting from the efficient CDF generation under user-specified stroke configurations, our method allows easy exploration of different PQ mesh layouts based on the same input shape, as shown in Fig.~\ref{fig:Layout_example_2}. More results on general models such as Stanford Bunny can be found in the supplementary materials.

\subsection{Computational Efficiency}

\begin{table}[t!]
\centering
    \begin{tabular}{c|c|c|c}
    \hline
    Model & \#Faces & Optimization & Ours  \\
    \hline
    Test Model 1 & 5000 & 2.851s & 0.200s (14.3$\times$) \\
    Test Model 2 & 5000 & 2.855s & 0.194s (14.7$\times$)  \\
    Test Model 3 & 5000 & 2.858s & 0.195s  (14.6$\times$)\\
    \hline    
    Vase & 23642 & 17.326s & 0.254s (68.2$\times$)  \\
    Dome & 44490 & 30.198s & 0.417s (72.4$\times$)  \\
    Face & 60077 & 40.412s & 0.571s (70.8$\times$)  \\
    \hline  
    Garden & 8322 & 4.946s & 0.206s (24.0$\times$)  \\
    Yas Island & 7029 & 3.766s & 0.204s (18.5$\times$)  \\
    Aqua Dome & 10790 & 6.522s & 0.217s (30.1$\times$) \\
    \hline        
    \end{tabular}
    \caption{Time cost comparison for CDF generation between the method in ~\cite{diamanti2014designing} and our method on the surfaces shown in Fig.~\ref{fig:Viewer_example}.}
    \label{tab:Time}
\end{table}

\begin{table*}[htb]
\centering
    \begin{tabular}{c|c|c|c|c|c|c}
    \hline
    \multirow{2}{*}{Model} & \multicolumn{2}{c|}{$\eta_{\text{mean}}$} 
    & \multicolumn{2}{c|}{$\eta_{\text{max}}$}
    & \multirow{2}{*}{$\delta$}
    & \multirow{2}{*}{$\theta$} \\
     \cline{2-5}  
    & before & after & before & after & & \\ 
    \hline
    Test Model 1 & 0.0062 & 0.0024 & 0.0425 & 0.0096 & $5.61^{\circ}$ & $8.96^{\circ}$ \\
    Test Model 2 & 0.0060 & 0.0025 & 0.0483 & 0.0123 & $10.76^{\circ}$ & $11.08^{\circ}$ \\
    Test Model 3 & 0.0065 & 0.0030 & 0.0566 & 0.0136 & $5.14^{\circ}$ & $8.99^{\circ}$ \\
    \hline    
    Vase & 0.0035 & 0.0015 & 0.0259 & 0.0058 & $5.32^{\circ}$ & $9.02^{\circ}$ \\
    Dome & 0.0108 & 0.0036 & 0.0399 & 0.0127 & $8.21^{\circ}$ & $12.04^{\circ}$ \\
    Face & 0.0162 & 0.0037 & 0.2098 & 0.0253 & $6.82^{\circ}$ & $19.74^{\circ}$ \\
    \hline  
    Garden & 0.0082 & 0.0028 & 0.1015 & 0.0168 & $7.21^{\circ}$ & $23.88^{\circ}$ \\
    Yas Island & 0.0151 & 0.0017 & 0.1845 & 0.0110 & $16.10^{\circ}$ & $22.55^{\circ}$ \\
    Aqua Dome & 0.0140 & 0.0032 & 0.4529 & 0.0203 & $17.70^{\circ}$ & $14.39^{\circ}$ \\
    \hline  
    Test Set & 0.0067 & 0.0023 & 0.0591 & 0.0118 & $8.31^{\circ}$ & $11.30^{\circ}$ \\
    \hline
    \end{tabular}
    \caption{Quantitative measurements on the PQ meshes shown in Fig.~\ref{fig:Viewer_example}. The average measurements on our test set are shown in the last row. The planarity before and after applying vertex perturbation optimization~\cite{deuss2015shapeop} are both listed. $\delta$, $\theta$ are measured in degrees.}
    \label{tab:Measurement}
\end{table*}

We also validate the computational efficiency of our learning-based CDF generation method by comparing with the traditional optimization-based method~\cite{diamanti2014designing}, which we also employed in our dataset generation pipeline (Section~\ref{seq:CDFGen}) and thus serves as the most direct baseline for comparison Tab.~\ref{tab:Time} shows the time differences on examples provided in Fig.~\ref{fig:Viewer_example}. 
Owing to the efficiency of the parallel inferences of the learned model, our method can generate CDF much faster (generally over an order of magnitude, and even close to two orders of magnitude for those cases with a larger number of faces). This feature particularly favors the real-world application that requires an iterative CDF design process.

\begin{figure}[t!]
    \includegraphics[width=\linewidth]{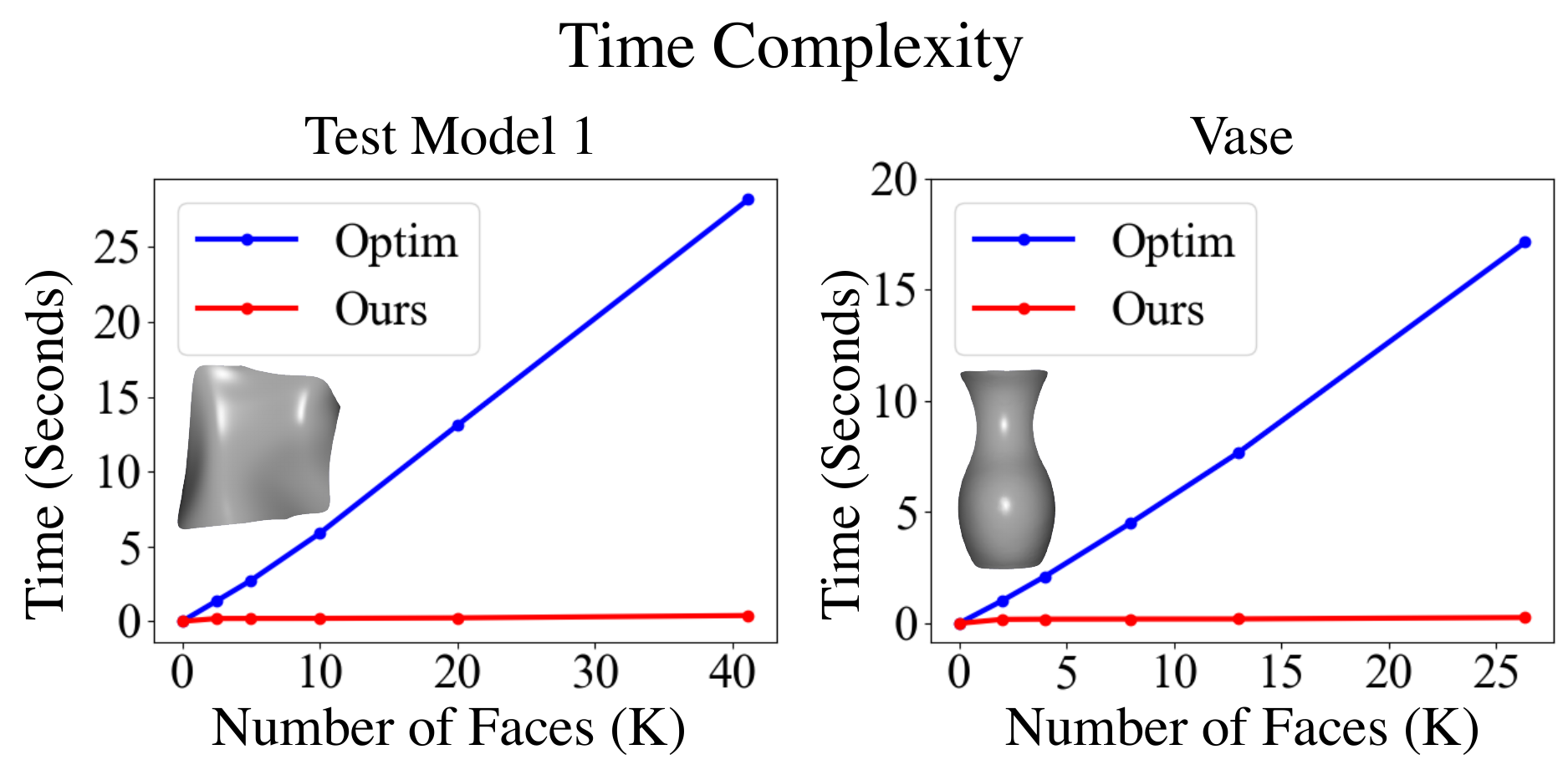}
	\caption{The time complexity illustration of the CDF generation between our method and the traditional optimization-based method~\cite{diamanti2014designing}.}
\label{fig:Time_Complexity}
\end{figure}

\begin{figure}[t!]
    \includegraphics[width=\linewidth]{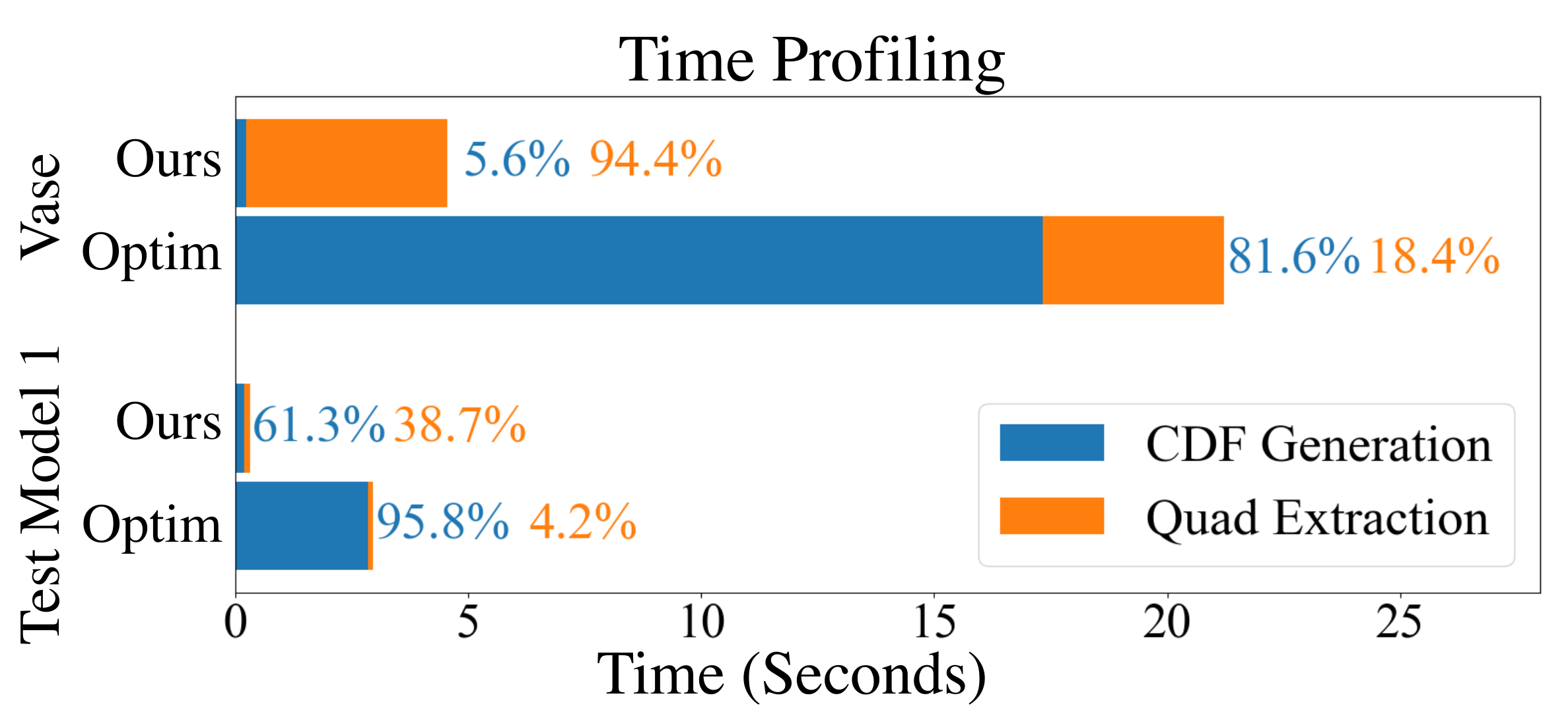}
	\caption{Time profiling of CDF generation and quad mesh extraction on our method and optimization method~\cite{diamanti2014designing}. The percentages are shown to the right.}
\label{fig:Time_Profiling}
\end{figure}

Additionally, Fig.~\ref{fig:Time_Complexity} illustrates the time complexity of our learning-based method over the traditional optimization-based method. 
For two test shapes with increased face numbers, we can see a clear growth (approximately linear) of the optimization cost for CDF generation. In contrast, due to the highly parallelized feedforward process of our learning-based approach, the time cost increases only marginally.
Fig.~\ref{fig:Time_Profiling} compares the cost of CDF generation and quad mesh extraction for two test models shown in Fig.~\ref{fig:Viewer_example}. We can see that for the optimization approach, CDF generation (even one round) takes the majority of the computational cost, and becomes a major bottleneck in the quad mesh design pipeline with iterative CDF attempts, especially for a bigger reference mesh such as Vase with $\sim$20k faces. On the contrary, our method can effectively address this issue and significantly reduce the time consumption, enabling a controllable and interactive CDF generation practice for user-friendly architectural design (see the supplementary video).

\subsection{Result Quality}
To evaluate the effectiveness of our method, we define a set of quantitative metrics to demonstrate the quality of our results.
First, we utilize the planarity of the quad mesh after quadrangulation to reflect the quality of the inferred CDF. For each quad face $f_j$, we compute the distance between two diagonals and divide it by the average edge length, resulting in $\eta_j$, where $\eta_j=0$ means $f_j$ is fully planar.
Then we obtain the maximum planarity error $\eta_{\text{max}}$ and the mean planarity error $\eta_{\text{mean}}$ for the entire quad mesh. 
We also evaluate the inferred CDF by directly measuring its consistency with the input strokes and closeness with the ground-truth CDF.
To make the metrics more geometrically meaningful, we use the angle between directions to measure the direction alignment instead of computing their inner-product as in Sec.~\ref{sec:Losses}.
For the stroke consistency, we define $\delta$  as the average angle between the inferred CDF and the strokes, where the directions are restricted on those mesh faces containing stroke segments. 
For CDF closeness, we define $\theta$ as the average angle between directions of the inferred CDF and the ground-truth CDF over all mesh faces.

Tab.~\ref{tab:Measurement} shows the quantitative evaluations on different models visualized in Fig.~\ref{fig:Viewer_example}, as well as the overall performance on our test set with 300 models.
The results show that our inferred CDFs are of high quality for guiding the quadrangulation, as the initial quad mesh layouts already exhibit good planarity. Also, the initial quad meshes serve as a good starting point for vertex perturbation optimization~\cite{deuss2015shapeop} where the planarity can be further improved.
The CDF closeness and consistency values also demonstrate that our method can directly infer CDFs adhering to the ground truth while respecting the stroke guidance.

\subsection{Ablation Study}
To validate the design choices of our method, we perform an ablation study on the loss functions defined in Sec.~\ref{sec:Losses}.
Note that the normal consistency loss $\mathcal{L}_{dn}$ is necessary for the inferred CDF directions to lie on the local tangent planes. Therefore, we only ablate the loss functions $\mathcal{L}_{ds}$ and $\mathcal{L}_{dc}$.
We measure the quality of the resulting CDF using $\delta$ and $\theta$ as defined in the previous section, as well as the number of singularities of the CDF. The comparison of the full method and the ablated methods are shown in Tab.~\ref{tab:Ablation}. 
We can see that with the help of loss function $\mathcal{L}_{ds}$, our result can produce far fewer singularities in the resulting PQ meshes. 
Furthermore, we can also generate a CDF with much better stroke consistency by adding loss $\mathcal{L}_{dc}$.
Some qualitative results are presented in the supplementary materials.

In addition, to evaluate the effectiveness of the feature representation of strokes described in Sec.~\ref{sec:FeatureRep}, we test an alternative approach where the strokes are simply treated as sequential data by themselves, and the feature is extracted from a Point Cloud Transformer (PCT)~\cite{guo2021pct}.
As shown in Tab.~\ref{tab:Ablation}, our method performs much better in all metrics, which verifies the superiority of our novel feature representation over PCT for surface strokes, as the surface context is also addressed in our representation.

\begin{table}[t]
\centering
    \begin{tabular}{c|c|c|c}
    \hline
     & \# of Sings & $\delta$  & $\theta$  \\
    \hline
     Without $\mathcal{L}_{ds}$ & 7.02 & $7.38^{\circ}$ & $10.55^{\circ}$  \\
    \hline
     Without $\mathcal{L}_{dc}$ & 4.67 & $10.32^{\circ}$ & $11.48^{\circ}$  \\
    \hline
     PCT~\cite{guo2021pct} & 8.97 & $20.98^{\circ}$ & $19.06^{\circ}$  \\
     \hline
     Ours & 4.91 & $8.31^{\circ}$ & $11.30^{\circ}$ \\
    \hline
    \end{tabular}
    \caption{Quantitative measurements of ablation studies. For the inferred CDFs on the test set, we measure the average number of singularities, $\delta$ - the consistency with the input strokes,  and $\theta$ - the closeness to the ground truth CDF. The ablated methods are without the direction smoothness loss $\mathcal{L}_{ds}$, without stroke consistency loss $\mathcal{L}_{dc}$, and without our context-aware stroke feature but using PCT~\cite{guo2021pct} for stroke feature extraction, respectively.}
    \label{tab:Ablation}
\end{table}

\begin{figure}[t]
    \includegraphics[width=\linewidth]{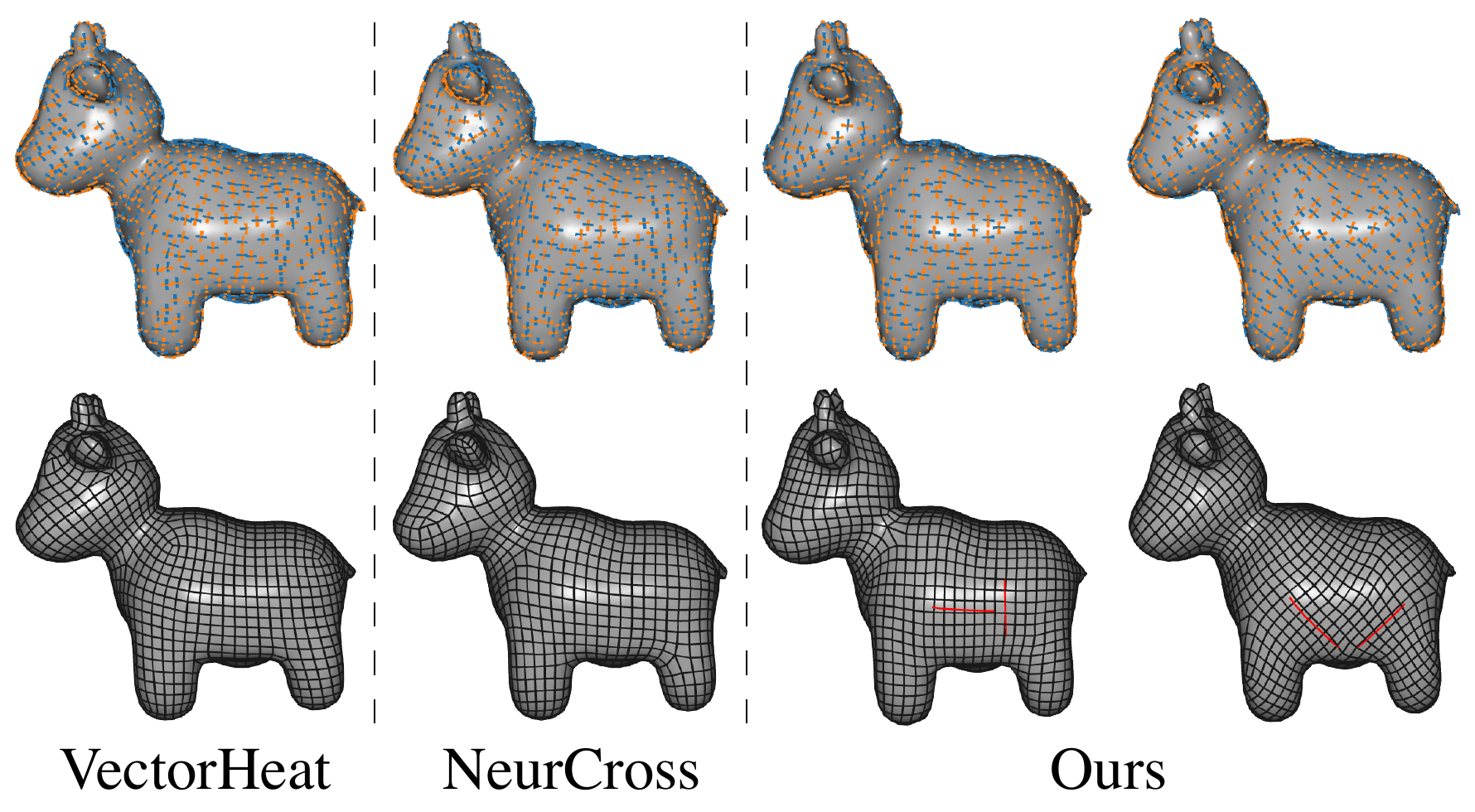}
	\caption{Comparison with recent learning-based field generation methods VectorHeat~\cite{gao2024vectorheatnet} and NeurCross~\cite{Dong2025NeurCross}. The generated direction field overlaid on the input reference surface (top row) and the resultant quad meshes (bottom row) are shown for each method. For clarity, the input strokes of our method (highlighted in red) are visualized on the final meshes.}
\label{fig:Comparison}
\end{figure}
 
\subsection{Comparison with Other Methods}

We also compare with recent learning-based field generation methods~\cite{gao2024vectorheatnet, Dong2025NeurCross} by contacting the authors and running their codes. The results are shown in Fig.~\ref{fig:Comparison}. VectorHeat~\cite{gao2024vectorheatnet} aims to generate rotation-invariant vector/cross fields. There is no guarantee that the generated cross field is a CDF, except for PDF which is a special type of CDF. Therefore, it cannot be used for PQ mesh optimization. NeurCross~\cite{Dong2025NeurCross} is restricted to PDF generation, thus cannot be used for controllable CDF generation as our method. In contrast, our method is capable of producing flexible CDF variations (and their corresponding PQ meshes) that align with the input strokes.

\section{Conclusion}
In this paper, we propose a novel learning-based method to infer conjugate direction fields for PQ mesh generation, avoiding complicated non-linear optimization.
Moreover, we design a unique feature representation for user-specified strokes that can be integrated with other geometry information for effective local and global feature encoding. 
We also present a synthetic dataset with a large number of freeform surfaces with ground truth CDFs and their guiding strokes, enabling the development of learning-based approaches.
Our work not only offers a much more efficient framework for CDF generation, but also contributes to intuitive control of the PQ mesh layout with user-specified strokes, largely benefiting interactive design applications in practice.

In the future, we would like to make improvements on surfaces with sharp features by introducing additional strokes along sharp edges.
Besides, while the direction smoothness loss term $\mathcal{L}_{ds}$ can eliminate or reduce the singularities on the output PQ mesh, there is no explicit control of the number and location of singularities, which is worth exploration. Finally, we plan to investigate the unsupervised approach for CDF generation with better generalization capabilities.

\section*{Acknowledgments}
We thank Zhi Deng for assistance with dataset preparation and Cherril Pope for proofreading the manuscript.
This work was supported by China Scholarship Council under Grant No. 202108340019.

\bibliography{aaai2026}

\section*{Appendix}

\begin{figure}[b!]
    \includegraphics[width=\linewidth]{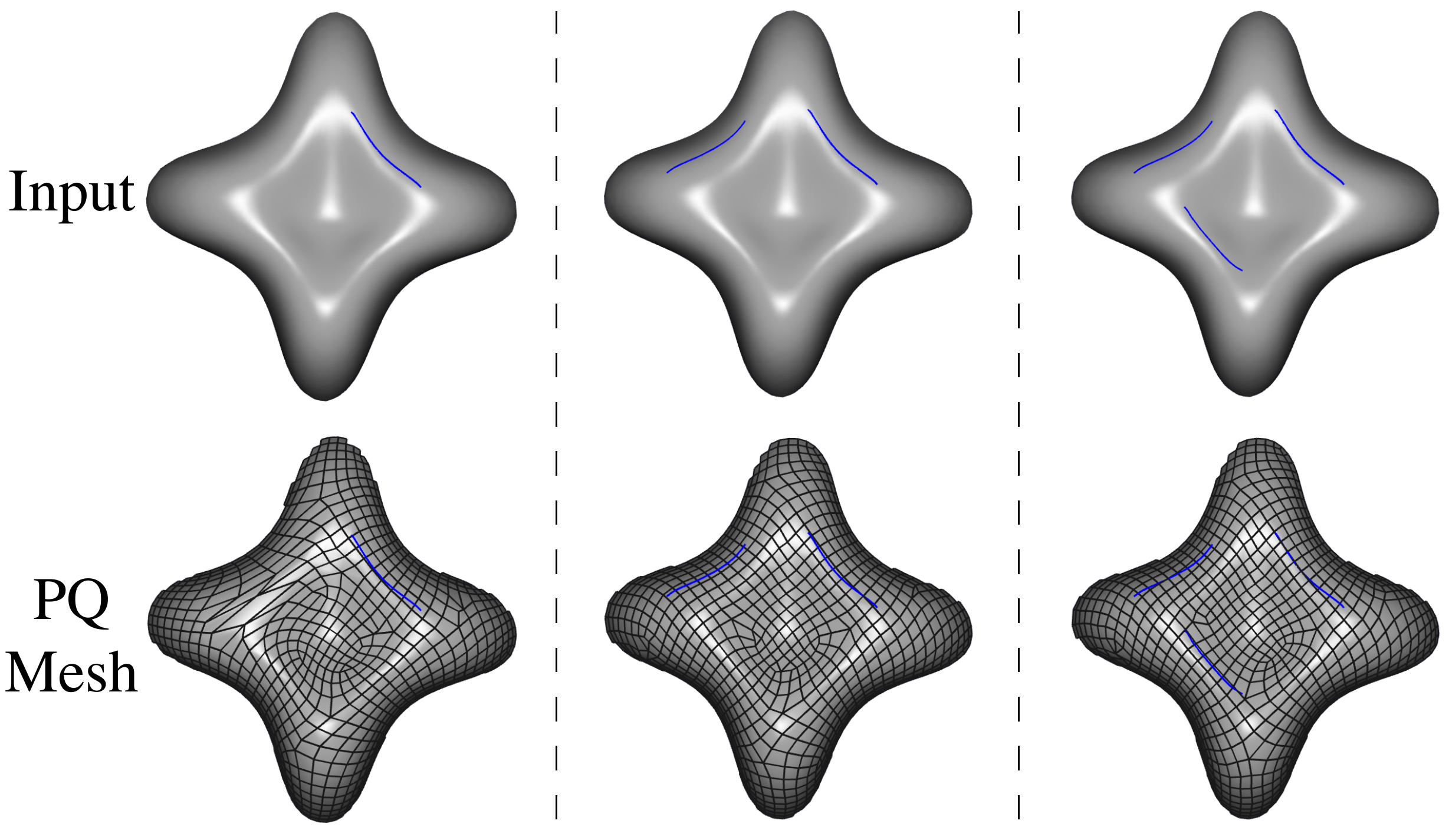}
	\caption{An example where the resulting PQ mesh layout is modified by
incrementally adding strokes.}
\label{fig:Layout_example_1}
\end{figure}

\begin{figure}[b!]
    \centering
    \includegraphics[width=.9\linewidth]{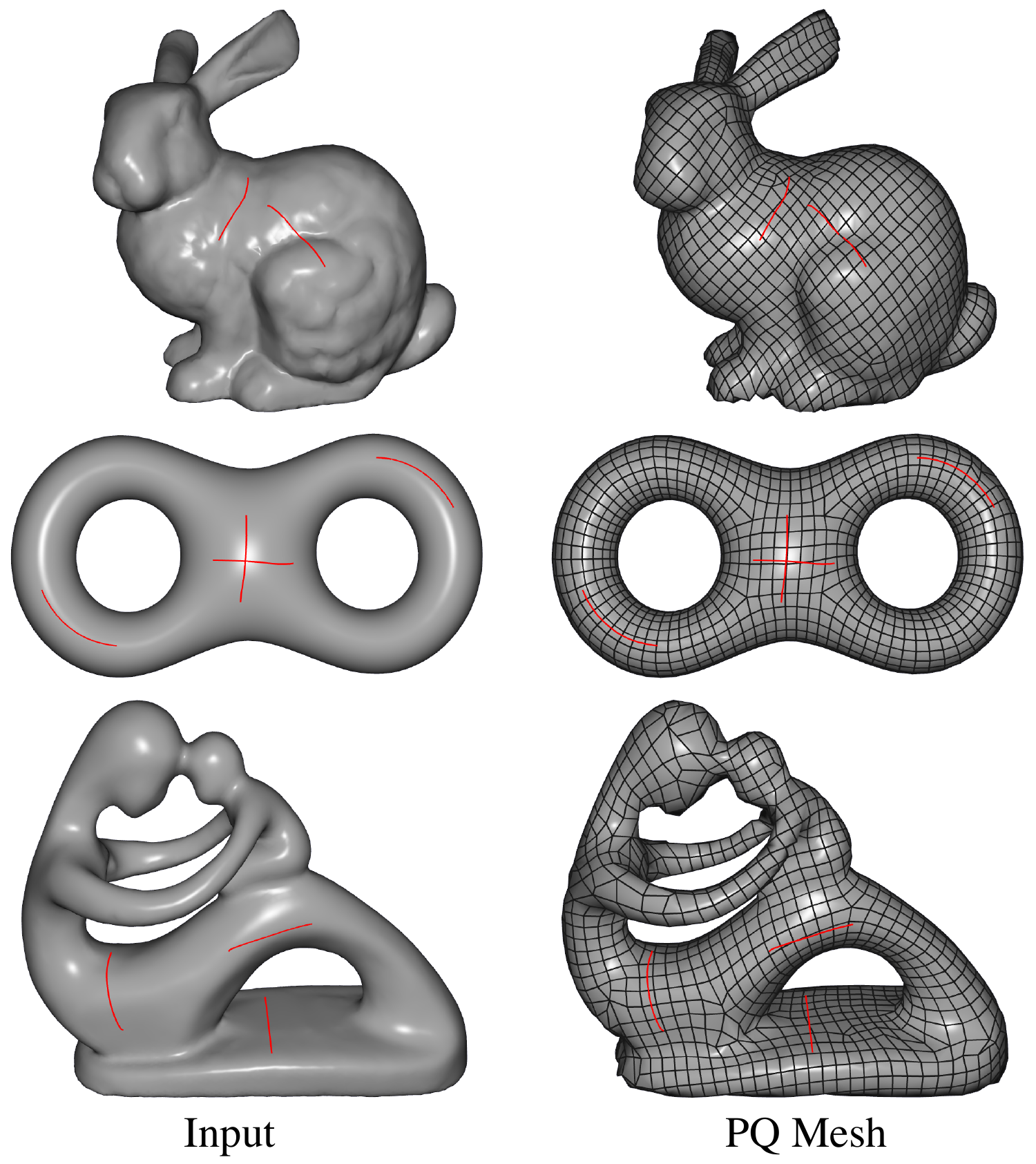}
	\caption{Examples of the PQ meshes generated using our approach on the general models. The red lines are user-specified strokes on the model. }
\label{fig:General_example}
\end{figure}

\begin{figure*}[t]
    \includegraphics[width=\textwidth]{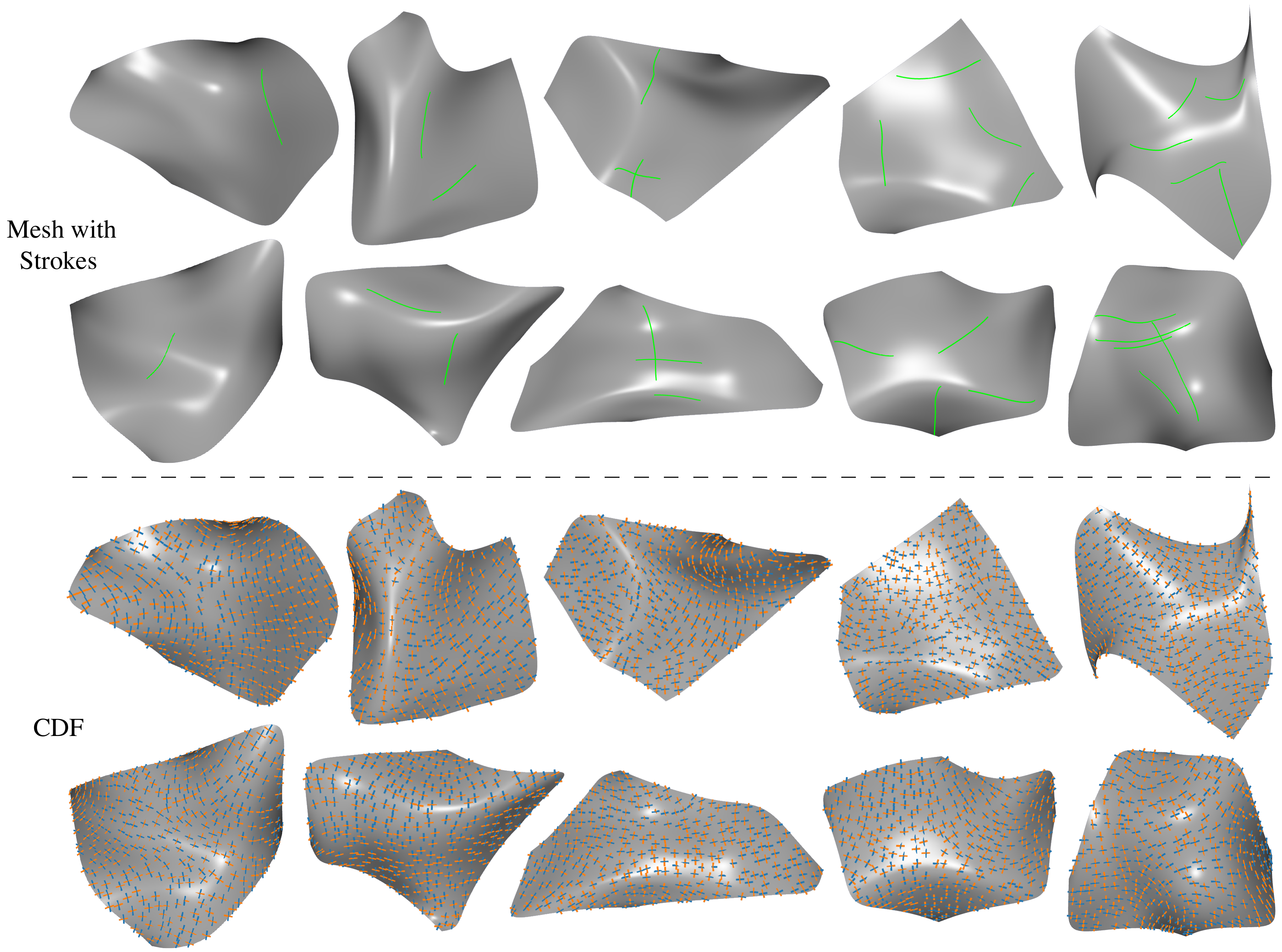}
	\caption{Data samples in the dataset generated using the method described in Section 4 of the main paper. The strokes (highlighted in green) on the freeform mesh surface (top two rows) are streamlines traced from the ground-truth CDF (bottom two rows).}
\label{fig:Dataset_example}
\end{figure*}

In this supplementary material, we further provide additional details and results to complement our main paper. More specifically, we present examples from our datasets, as well as other results of our proposed method, including PQ mesh layout control based on user-guided conjugate direction field (CDF) generation, qualitative results of the ablation study, and evaluations on general models such as Stanford Bunny.

\subsection*{Dataset Examples}
To demonstrate the characteristics of our dataset, we show a set of data samples from our dataset (see Fig.~\ref{fig:Dataset_example}), including the freefrom mesh surfaces with their corresponding streamlines as stroke mimics, as well as the ground truth CDFs.

\subsection*{Extended Layout Control}
 In our main paper (Figs. 6 and 7 therein), we show that our proposed method has the capacity to generate different PQ mesh layouts based on varying input strokes. Here we show that the PQ mesh layout can be easily controlled and updated by adding additional strokes for CDF generation, fulfilling real-world design requirements (see Fig.~\ref{fig:Layout_example_1}). 

\subsection*{Qualitative Results of Ablation Study}
In addition to the quantitative evaluation in Tab. 3 of the main paper, here we show some qualitative results of our ablation study in Fig.~\ref{fig:Ablation_example}. It can be seen that by leveraging the loss term $\mathcal{L}_{ds}$, our method effectively reduces the number of singularities in the generated PQ meshes. Also, adding the loss term $\mathcal{L}_{dc}$ allows better alignment of the CDF with the input strokes. 

\begin{figure}[t!]
    \includegraphics[width=\linewidth]{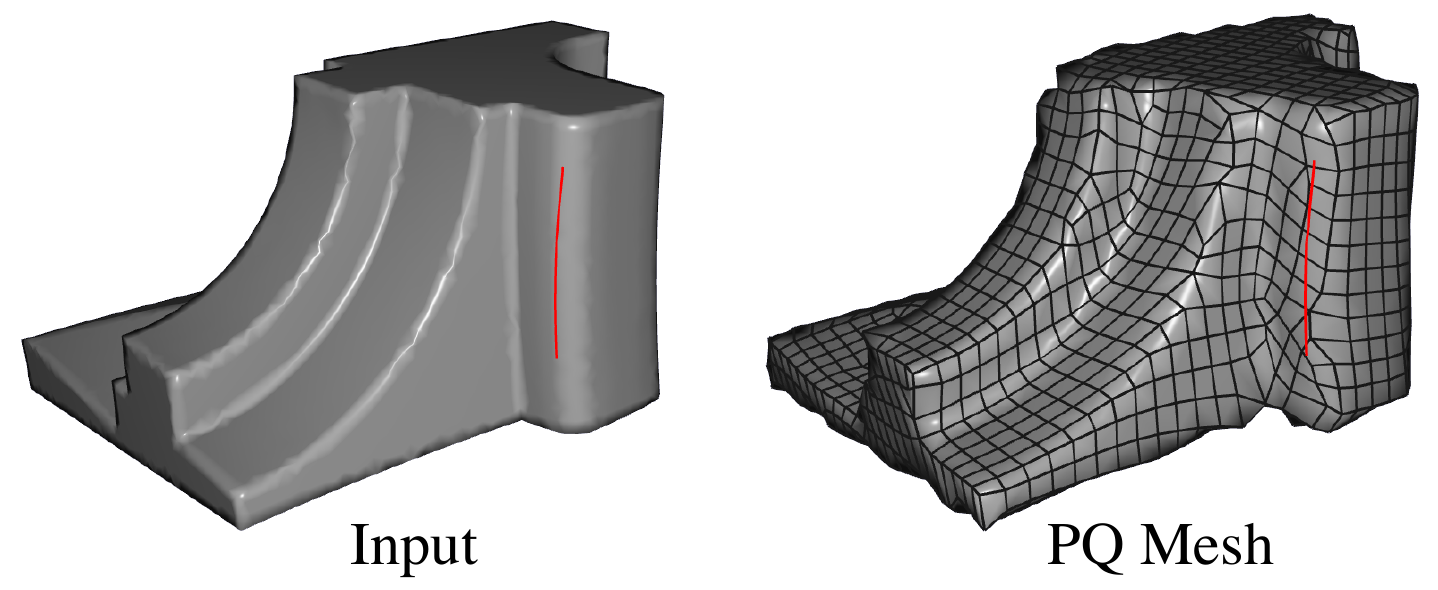}
	\caption{Example of the PQ mesh generated using our approach on the model with sharp features. The red line is user-specified stroke on the model.}
\label{fig:Failure_example}
\end{figure}

\begin{figure*}[t]
    \includegraphics[width=\textwidth]{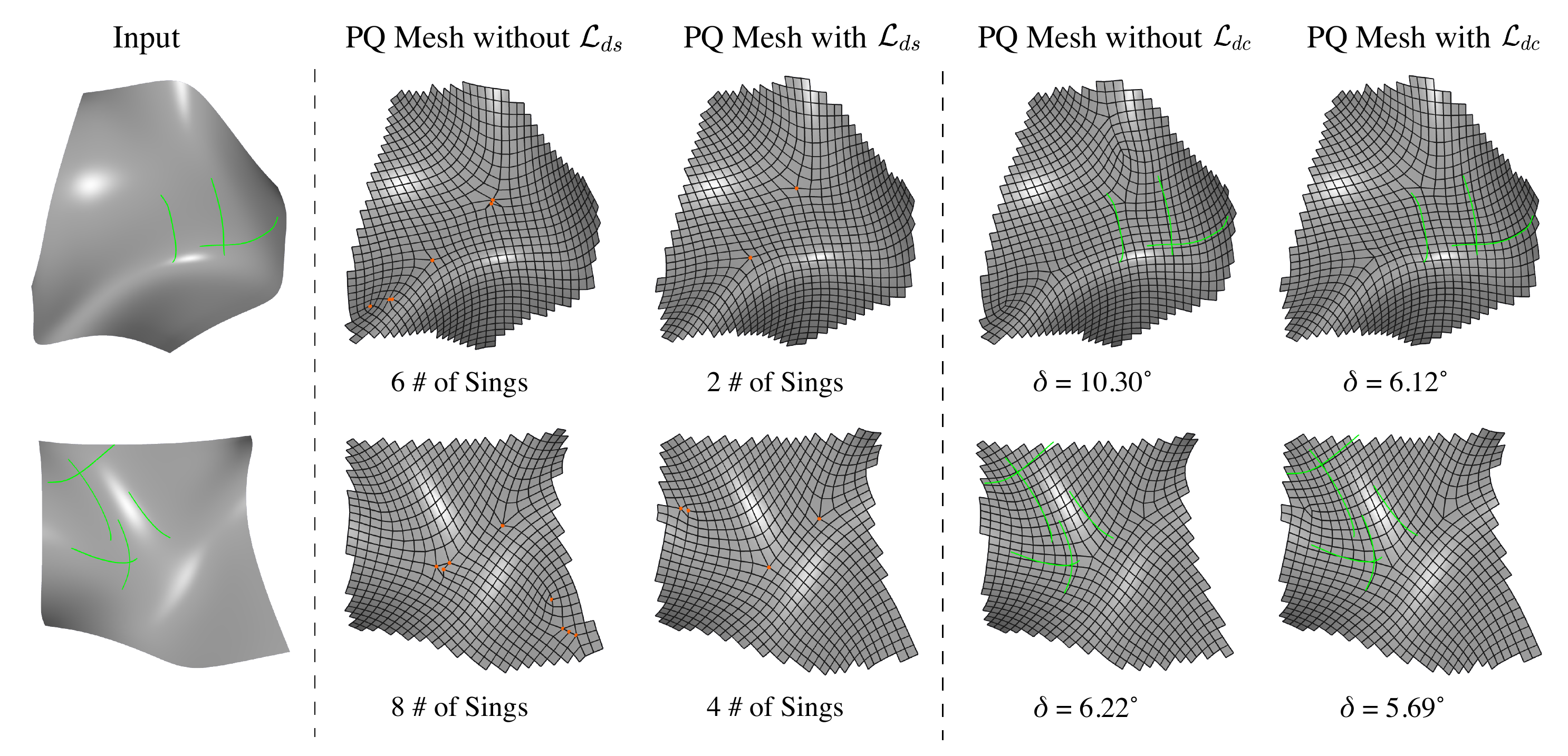}
	\caption{Qualitative results by ablating loss terms $\mathcal{L}_{ds}$ and $\mathcal{L}_{dc}$ for CDF generation. For clarity, we show the final PQ mesh instead of the CDF that controls mesh edge layout. We also visualize singular points in the generated PQ mesh as red points, where ``\# of Sings" denotes the number of singular points.}
\label{fig:Ablation_example}
\end{figure*}

\subsection*{Further Experiment on General Models}
Although we utilize open-boundary B-spline surfaces to construct our training and test dataset, and our method works on various open-boundary surfaces as demonstrated in our main paper, we further explore the generalization capacity of our method on closed models. Fig.~\ref{fig:General_example} shows some inspiring examples with different topologies. 
The layout of the resulting PQ meshes well aligns the input strokes in all examples.

On the other hand, as shown in Fig.~\ref{fig:Failure_example}, our inferred CDF given very few strokes fails to accurately align the sharp features. This is because we mainly use strokes to guide smooth CDF generation, while the sharp features are not taken into account in the dataset and the training process.



\end{document}